%% file: 0a.main.tex
\documentclass[10pt,conference]{IEEEtran}
\IEEEoverridecommandlockouts
\input{0b.preamble}
\input{1a.authors}		
\makeatletter
\patchcmd{\maketitle}{\@fnsymbol}{\@alph}{}{}  
\makeatother					
\title{Quality-Aware Task Offloading for Cooperative Perception in Vehicular Edge Computing}
\begin{document}
	\bstctlcite{IEEEexample:BSTcontrol}
\maketitle

	\input{1b.abstract}
	\input{2a.introduction}
	\input{2b.related_works}

\input{2c.system_model}

	\input{2d.problem_formulation}

	\input{2e.experiments}

	\input{2e.conclusion}

	\input{3b.references}
\end{document}

%% file: 0b.preamble.tex
\usepackage{cite}
\usepackage{amsmath,amssymb,amsfonts}
\usepackage{color,soul}
\usepackage{graphicx}
\usepackage{textcomp}
\usepackage{relsize}
\usepackage{bm}
\usepackage{booktabs}
\usepackage[table]{xcolor}
\usepackage{pgfplotstable}
\usepackage{colortbl}
\usepackage{pgfplots}
\usepackage{stfloats}
\usepackage{cuted}
\usepackage{caption}
\usepackage{marginnote}
\usepackage{verbatim}
\usepackage[table]{xcolor}
\usepackage{mathtools}
\usepackage[makeroom]{cancel}
\usepackage{enumitem}
\usepackage{pifont}
\usepackage{tikz}
\usepackage{relsize}
\usepackage{pgfplots}
\usetikzlibrary{shapes, backgrounds, arrows}
\usepackage{dsfont}
\usepackage{bbm} 
\usepackage{stmaryrd} 
\usepackage[numbered,framed]{matlab-prettifier}
\usepackage{bbold}
\usepackage{subcaption}
\usepackage{longtable}
\usepackage{algpseudocode}
\usepackage{algorithm}
\usepackage{ifthen}
\usepackage{caption}
\usepackage{subcaption}
\usepackage{eqparbox}
\usepackage{blkarray}
\usepackage{amsmath}
\usepackage[short]{optidef}
\usepackage{mathtools,graphicx}
\usepackage{etoolbox}
\usepackage{orcidlink}

\algnewcommand{\LeftComment}[1]{\Statex \(\triangleright\) #1}

\def\thesubsubsectiondis{\unskip\arabic{subsubsection})}
\usepackage{amsthm}
\allowdisplaybreaks

\graphicspath{ {./Figures/} }
\bstctlcite{IEEEexample:BSTcontrol}

\makeatletter
\newcommand\fs@norules{\def\@fs@cfont{\bfseries}\let\@fs@capt\floatc@ruled
	\def\@fs@pre{}%
	\def\@fs@post{}%
	\def\@fs@mid{\kern3pt}%
	\let\@fs@iftopcapt\iftrue}
\newcommand{\linebreakand}{%
\end{@IEEEauthorhalign}
\hfill\mbox{}\par
\mbox{}\hfill\begin{@IEEEauthorhalign}
}

\makeatother

\DeclareSymbolFont{extraup}{U}{zavm}{m}{n}
\DeclareMathSymbol{\varheart}{\mathalpha}{extraup}{86}
\DeclareMathSymbol{\vardiamond}{\mathalpha}{extraup}{87}

\definecolor{mycolor}{rgb}{0.8,0.3,0}
\newlist{todolist}{itemize}{2}
\setlist[todolist]{label=$\square$}						

%% file: 1a.authors.tex
\author{
	Amr M. Zaki, 
	Sara A. Elsayed, 
	Khalid Elgazzar,
	Hossam S. Hassanein \textsuperscript{\orcidlink{0000-0003-0260-8979}}
	\thanks{Amr M. Zaki (corresponding author), Sara A. Elsayed and Hossam S. Hassanein are with Queen’s University, Kingston, ON K7L 2N8,
		Canada  (e-mail: amr.zaki@queensu.ca; selsayed@cs.queensu.ca; hossam@cs.queensu.ca) . Khalid Elgazzar is with Ontario Tech University, Oshawa, ON L1G 0C5, Canada (e-mail: khalid.elgazzar@ontariotechu.ca).}
}

%% file: 1b.abstract.tex
\begin{abstract}
	
	Task offloading in Vehicular Edge Computing (VEC) can advance cooperative perception (CP) to improve traffic awareness in Autonomous Vehicles.
	In this paper, we propose the Quality-aware Cooperative Perception Task Offloading (Q-CPTO) scheme. Q-CPTO is the first task offloading scheme that enhances traffic awareness by prioritizing the quality rather than the quantity of cooperative perception. Q-CPTO improves the quality of CP by curtailing perception redundancy and increasing the Value of Information (VOI) procured by each user. We use Kalman filters (KFs) for VOI assessment, predicting the next movement of each vehicle to estimate its region of interest. The estimated VOI is then integrated into the task offloading problem. We formulate the task offloading problem as an Integer Linear Program (ILP) that maximizes the VOI of users and reduces perception redundancy by leveraging the spatially diverse fields of view (FOVs) of vehicles, while adhering to strict latency requirements. We also propose the Q-CPTO-Heuristic (Q-CPTO-H) scheme to solve the task offloading problem in a time-efficient manner. 
	Extensive evaluations show that Q-CPTO significantly outperforms prominent task offloading schemes by up to 14\% and 20\% in terms of response delay and traffic awareness, respectively. Furthermore, Q-CPTO-H closely approaches the optimal solution, with marginal gaps of up to 1.4\% and 2.1\% in terms of traffic awareness and the number of collaborating users, respectively, while reducing the runtime by up to 84\%.

\end{abstract}

%% file: 2a.introduction.tex
\section{Introduction}
With the advent of Autonomous Vehicles, significant societal benefits can be garnered, including a substantial decrease in traffic accidents and fatalities \cite{800billion}. vehicles rely on various on-board sensors (LiDAR, cameras, radar, GPS) to perceive their surroundings and ensure safe operation in complex traffic environments \cite{Vehicular-Edge-Computing-intro,Accurate-representation-core-function}. However, relying solely on the data collected from a single vehicle can introduce significant vulnerabilities, due to sensor impairments. These impairments can include occlusions from static or dynamic obstacles, such as buildings or other vehicles, and limited perception horizons due to the restricted field of view (FOV) and low-point density in distant regions \cite{coop_3d},\cite{Percpetion-task-2023}. Consequently, the reliability of an individual vehicle’s perception can be compromised, potentially jeopardizing road safety and traffic efficiency \cite{CoopPerception-2021-Decentralized}.

Leveraging Cooperative Perception (CP) can alleviate the aforementioned problems \cite{CoFF},\cite{coop_3d},\cite{4-Terabytes}. CP involves the fusion of sensor data from multiple vehicles, thus augmenting the detection capabilities of individual vehicles and enhancing the overall traffic situational awareness \cite{Vehicular-Edge-Computing-intro}. However, CP imposes substantial demands on the vehicle’s on-board computing resources, due to its time-critical and data-intensive nature \cite{4-Terabytes}. Vehicular Edge Computing (VEC) is an auspicious computing paradigm that can foster such demands to satisfy the stringent Quality of Service (QoS) requirements associated with CP \cite{coop_3d}\cite{fog-follow-me-cloud-delay}. 

In VEC, Roadside Units (RSUs) and VEC servers are used as edge nodes by exploiting their increasingly powerful on-board computing units \cite{fog-follow-me-cloud-delay}. Additionally, VEC brings the computing service closer to vehicles and end-users, significantly reducing latency and curtailing ineffective transmissions \cite{sec2021}. Consequently, task offloading in VEC can facilitate CP by enabling each vehicle to offload its CP task to an edge node, which receives data frames from diverse vehicles, aggregates them, and then sends the fused result back to the requesting vehicles \cite{Case_Cooperative_Perception_globecomm2020}. To leverage CP effectively, efficient task offloading strategies are crucial.

Existing task offloading schemes focus mostly on optimizing certain QoS metrics, including latency and energy consumption \cite{ICC2023-CPTO,VEC24-taskoffloading-delay-energy-1,VEC24-taskoffloading-delay-energy-2,VEC24-taskoffloading-delay-energy-3,VEC24-taskoffloading-delay-energy-attention-withtransformenrs}. Other schemes account for the potential risk of delivery failure due to path loss, while abiding by the strict QoS requirements of cooperative perception \cite{PLTO}. Recently, some attempts have been made to improve situational awareness by increasing the number of collaborating vehicles, while maintaining a certain QoS \cite{ICC2023-CPTO}. However, such schemes focus on the quantity rather than the quality of cooperative perception. For instance, they fail to mitigate data duplication and redundancy issues, where in congested urban scenarios, multiple vehicles may possess overlapping views and perceptions, transmitting duplicate data about the same objects \cite{CoopPerception-2020-Redundancy}. Note that it has been shown that vehicles can send identical data about detected objects as much as 25 to 50 times per second \cite{CoopPerception-2020-Redundancy}. Furthermore, existing schemes overlook individual user preferences for acquiring additional perception in specific road areas, disregarding the possibility that certain regions might hold no interest for them. Failing to account for such factors can profoundly impact the quality of CP, significantly limiting situational awareness.

In this paper, we propose the Quality-aware Cooperative Perception Task Offloading (Q-CPTO) scheme. Q-CPTO improves the quality of cooperative perception by increasing the Value of Information (VOI) gained by each individual user and mitigating perception redundancy. The VOI of each user (i.e., requesting vehicle) is estimated based on its next intended movement. This is since additional perceptions gained in a certain direction tend to be more valuable for vehicles turning in that same direction. Accordingly, Q-CPTO predicts the next movement of each vehicle to estimate the vehicle’s Region of Interest (ROI). We make task offloading decisions that determine the collaborating vehicles associated with each edge node by considering the vehicles’ shared ROI to optimize the quality of CP. We also leverage the spatial distribution of the vehicles’ Field-of-View (FOV) to decrease redundancy. We formulate the task offloading problem as an Integer Linear Program (ILP). Our contributions can be summarized as follows:
\begin{itemize}[]
	\item We propose a novel task offloading scheme (Q-CPTO) that maximizes situational awareness by accounting for the quality of cooperative perception. Q-CPTO is the first scheme that maximizes the VOI of each user and curtails the redundancy of received perceptions, while abiding by a certain deadline to adhere to the strict delay requirements of CP.
	
	\item We interweave the notion of VOI with CP. In particular, we identify the VOI of each user based on its personalized interest in the additional perception gained by its potential collaboration with each vehicle. Towards that end, we estimate the VOI of each user by predicting its next turn intention at an intersection using Kalman Filtering (KF).
	
	\item We reduce the runtime by introducing the Q-CPTO-Heuristic (Q-CPTO-H) scheme to solve the task offloading problem in a time-efficient manner.
\end{itemize}
 
Extensive evaluations show that Q-CPTO achieves significant improvements in terms of response delay and situational awareness compared to prominent task offloading schemes. Additionally, Q-CPTO-H comes remarkably close to the optimal solution, exhibiting slight disparities in terms of situational awareness and the number of collaborating users, while significantly reducing the runtime. 

The remainder of the paper is organized as follows. Section \ref{sec:related_works} highlights some of the related work. Section  \ref{sec:System Model} presents the proposed schemes (Q-CPTO and Q-CPTO-H). Section \ref{sec:exps} discusses the performance evaluation and simulation results. Section \ref{sec:conclustions} provides concluding remarks and draws future directions.

%% file: 2b.related_works.tex
\section{Related Work}
\label{sec:related_works}

In recent years, the offloading of vehicular computing tasks to the edge has emerged as a crucial area of research to ensure the safe and efficient operation of vehicles \cite{ICC2023-CPTO,VEC24-taskoffloading-delay-energy-1,VEC24-taskoffloading-delay-energy-2,VEC24-taskoffloading-delay-energy-3,VEC24-taskoffloading-delay-energy-attention-withtransformenrs}.  This shift has been motivated by the computation-intensive nature and stringent delay requirements associated with these tasks. Various solutions have been proposed to address the challenge of reducing offloading latency.

Feng et al. \cite{VEC24-taskoffloading-delay-energy-1} focused on the joint optimization of task partitioning ratios and user association, formulating it as a mixed-integer programming problem with the specific goal of minimizing average latency for all users. Chen et al. \cite{VEC24-taskoffloading-delay-energy-2} tackled the complexities of optimizing task execution delay by treating it as a min–max problem, employing a particle swarm optimization-based approach. Ke et al. \cite{VEC24-taskoffloading-delay-energy-3} developed a task offloading algorithm tailored for VEC environments, aiming to strike a balance between energy use and data transfer latency.
However, these studies primarily focused on optimizing the offloading decision, neglecting the heterogeneity in VEC environments among multiple tasks. To address this gap, Gao et al. \cite{VEC24-taskoffloading-delay-energy-attention-withtransformenrs} introduced a decentralized task offloading method utilizing transformers and a policy decoupling-based multi-agent actor–critic (TPDMAAC) framework. In their model, each vehicle independently selects the optimal worker and allocates necessary resources to optimize both task latency and energy consumption, thereby enhancing system efficiency.
In a different context, Zaki et al. \cite{PLTO} proposed a scheme that concurrently optimizes task offloading with respect to both delay and the number of cooperating users to workers with a low probability of path loss.

Some works have also aimed at enhancing the perception capabilities of vehicles through task offloading. For instance, Krijestorac \cite{Case_Cooperative_Perception_globecomm2020} proposed a task offloading scheme improving cooperative perception by maximizing the data transmission rate of vehicles. Another study by Zaki et al. \cite{ICC2023-CPTO} introduced the Cooperative Perception-based Task Offloading (CPTO) scheme. This scheme minimizes the latency of perception aggregation tasks at the worker and maximizes the number of served requests, thereby enhancing the situational awareness of vehicles through an increased number of cooperating vehicles.
 Despite these advancements, existing research has predominantly focused on augmenting the cooperation aspect by increasing the number of participating vehicles.  However, crucial aspects influencing perception quality, including the spatial-temporal distribution of vehicles, redundancy in perception, and the practical benefits derived from received perceptions, have been overlooked.

Therefore, it is imperative, yet challenging, to study the quality of cooperative perception from the users' perspective while adhering to the stringent latency constraints of perception tasks. 
This paper marks the first effort to optimize the VOI of offloaded tasks by leveraging the shared interest between vehicles while adhering to the latency restrictions of CP.

%% file: 2c.system_model.tex
\section{QUALITY-AWARE COOPERATIVE PERCEPTION TASK OFFLOADING (Q-CPTO)}
\label{sec:System Model}





\begin{figure*}
	\centering
	\includegraphics[width=\linewidth]{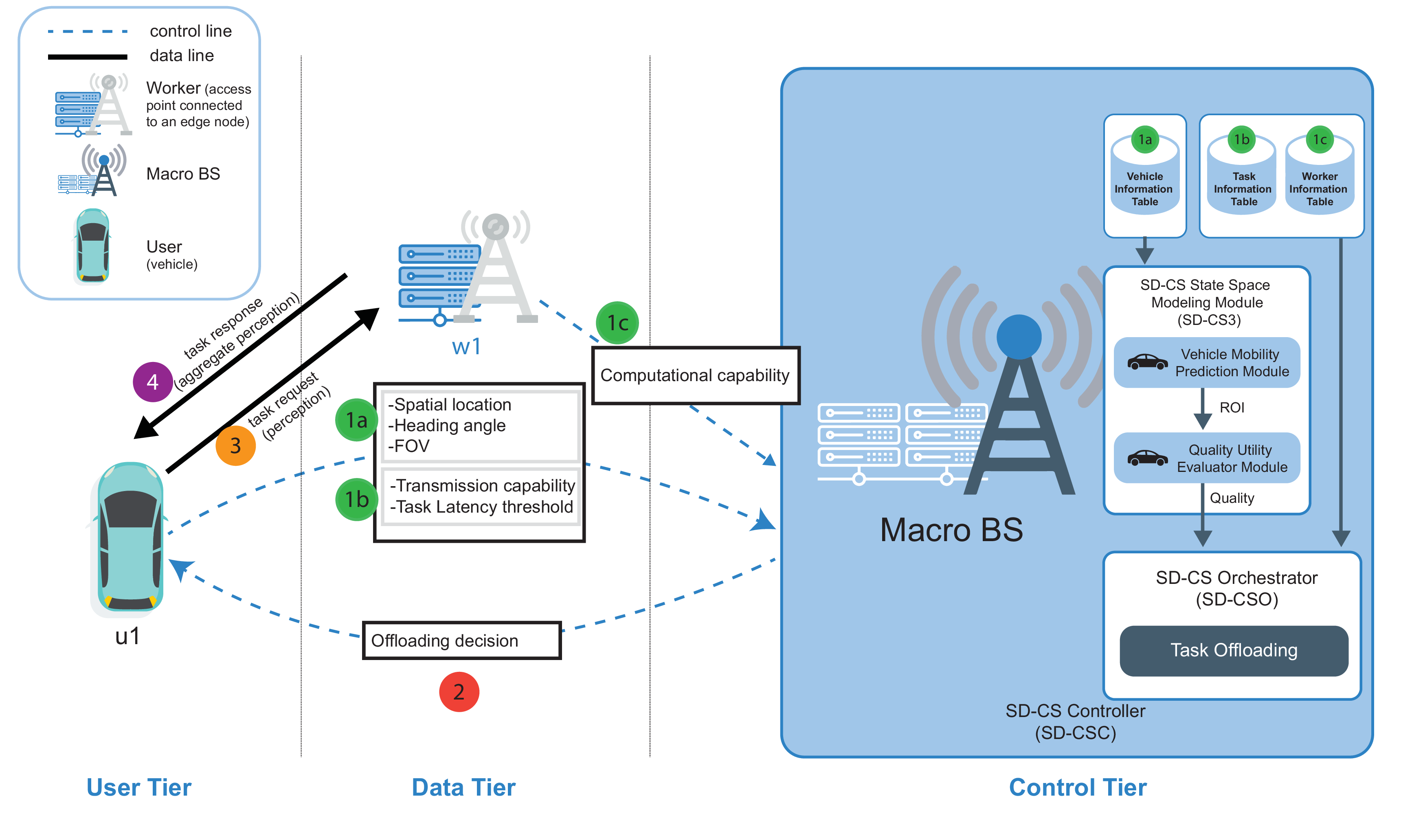}
	\caption{Software-defined network for cooperative services (SD-CS) framework.}
	\label{fig:system-model}
\end{figure*}
\subsection{System Architecture}
Q-CPTO uses a centralized framework that adopts the same system architecture proposed by Chen et al. \cite{SDN-paper}. We refer to the underlying framework as the Software-Defined network for Cooperative Services (SD-CS). As depicted in Figure \ref{fig:system-model}, SD-CS encompasses three tiers: user, data, and control.
The user tier consists of the requesting vehicles (i.e., users) subscribing to the system to offload their perception tasks to VEC servers (i.e., workers).
The latter represents the data tier, which corresponds to Roadside units (RSUs) that serve as access points to edge nodes.
The control tier contains the SD-CS controller (SD-CSC), which is the centralized entity responsible for making all vital decisions, including the prediction and task offloading decisions. This controller includes two modules; the SD-CS state space modeling module (SD-CS3) and the SD-CS orchestrator module (SD-CSO). 

SD-CSC is  deployed on a macro cell Base Station (BS). Wireless communications enables user interactions with both workers and the macro BS.

The control functionalities, including the task offloading policy, are centrally managed by the SD-CSC. As depicted in Figure \ref{fig:system-model}, users periodically transmit essential data, including wireless transmission capability, current locations, heading angles, field-of-view (FOV) and task latency threshold requirements to the SD-CSC (Steps 1a,1b).
Similarly, RSUs relay information about their computational capabilities to the SD-CSC (Step 1c).
The SD-CSC organizes this incoming data into information tables, comprising a user information table, a BS information table, and a task information table.
Upon receiving information about a task request, the SD-CSC initiates a series of actions. This includes the execution of the ROI estimation module (i.e., mobility prediction module) and Quality Utility Evaluator Module within SD-CS3  (elaborated upon in subsequent sections). Subsequently, the SD-CSO executes the task offloading policy.

Once users receive the task offloading decisions from the SD-CSO (Step 2), they offload their CP tasks to the designated workers (Step 3). The workers then apply CP fusion algorithms to the received perceptions, the fusion result is then disseminated to the served users, thereby enhancing their situational awareness (Step 4).

\subsection{System Model and Overview }
Consider a set of users $U=\{u_1, u_2,..., u_{n}\}$ and a set of worker $W =\{w_{1}, w_{2},..., w_{m}\}$, where time is discretized into slots of equal duration $\Delta$, represented by $\mathcal{T} = \{1, \ldots ,T\}$.
In each time slot $\tau\in \mathcal{T}$, each user $u_i \in U$ transmits its CP task to the designated worker $w_j$. 
We adopt a widely used task model \cite{task-profile} to characterize the perception task, denoted as $\Psi_i = \{l_i,\lambda_i,\kappa_i\}$
,where $l_i$ represents the computation workload (in CPU cycles), $\lambda_i$  constitutes the size of the perception frame (in bits), and $\kappa_i$ represents the latency threshold (in msec). 

Each worker $w_j \in W$ has a maximum CPU frequency, denoted $C_j$ (in CPU cycles/sec), which is communicated to the SD-CSC. 
We assume that the computational resources of each worker is divided equally among the collaborating users assigned to it \cite{fog-follow-me-cloud-delay}. 
The number of tasks offloaded to each worker $w_j$ corresponds to the number of vehicles concurrently utilizing this worker, denoted $\eta_j$ (i.e., collaborating users).
The data rate between user $u_i$ and worker $w_j$ is denoted $R_{ij}$ in time slot $\tau \in \mathcal{T}$ \cite{79-peer-offloading}.

The response latency of the
perception task of user $u_i$ on worker $w_j$ is denoted $t_{ij}$, and is given by Eq.  \ref{eq:total_latency}, where $\alpha_{ij}$ is the computation delay and $\gamma_{ij}$ is the transmission latency. Since we focus on edge
computing systems, the users and the workers are in close vicinity; therefore, the propagation delay is negligible \cite{VEC24-taskoffloading-resourceallocation-SDP}.

\begin{equation}
	t_{ij} = 	\alpha_{ij} + \gamma_{ij}
	\label{eq:total_latency}
\end{equation}

The computation delay $\alpha_{ij}$ is the time it takes for worker $w_j$ to execute the perception aggregation task of user $u_i$, as given by Eq. \ref{eq:total_latency}.
\begin{equation}
	\alpha_{ij} = 	 \frac{l_i }{C_j/ \eta_j}
	\label{eq:cpu_latency}
\end{equation}
 The transmission delay $\gamma_{ij}$ represents the time required to transmit the entire perception frame from user $u_i$ to worker $w_j$, as given by Eq. \ref{eq:cpu_latency}. Note that the transmission delay associated with sending the result back to the user is disregarded, since its size is negligible \cite{Case_Cooperative_Perception_globecomm2020}.

\begin{equation}
	\gamma_{ij} = 	 \frac{\lambda_i}{R_{ij}}
	\label{eq:tr_latency}
\end{equation}

\subsection{Vehicle Mobility Prediction Module}

In order to estimate the VOI of each user, Q-CPTO predicts the next movement of each vehicle. The prediction process is conducted by the SD-CS3.
SD-CS3 considers the vehicle's current location and heading direction (from the vehicle information table) to predict its next turn.
This prediction is used to estimate the vehicle's ROI, which represents the area where the vehicle is likely to move in the next time slot $(\tau+1) \in \mathcal{T}$. We use Kalman filtering (KF) \cite{kalman-details-1} for such predictions.

Kalman filter (KF) is an effective recursive filter that estimates the state of a linear dynamic system
using a series of noisy measurements \cite{kalman-details-1}\cite{kalman-extended-version-1}. It operates using a set of mathematical equations and state space models, functioning as a predictor-corrector type estimator. The primary objective of KF is to optimally solve these mathematical equations, with the aim of minimizing the estimated error covariance. The Kalman filter's estimation prowess is particularly vital in accurately determining the motion state of interest.

The motion state of interest is represented as $X$, which signifies the 2D position of a vehicle. It serves as the essential data for characterizing the dynamic behavior of the vehicular system. Specifically, $X$ is the minimal amount of historical 2D positions that KF requires to predict future next location accurately.

This prediction is intricately reliant on several matrices and noise terms, namely the state transition and measurement matrices, denoted $A$ and $H$, respectively. The state transition matrix represents how the vehicle's state (2D location) evolves over time. It encapsulates the dynamics of the vehicle's motion, predicting its future state based on its current state and any known inputs or influences. The measurement matrix relates the true state of the vehicle to the measurements obtained from sensors or observations. It establishes the connection between the vehicle's actual state and the data collected from various sensors, aiding in refining the estimation of the vehicle's state.

These matrices, $A$ and $H$, work in tandem with the disturbance input matrix and sensor measurements, denoted $B$ and $Z$, respectively. The disturbance input matrix $B$ represents the influence of external factors or control inputs on the vehicle's motion. These external factors captured by $B$ could include forces, inputs, or disturbances that affect the vehicle's dynamics but are not directly measured or observed by the system.

The matrices $B$ and $Z$, along with the process noise and measurement noise, denoted  $\omega_{\tau}$ and $\upsilon_{\tau}$, respectively, encompass both external influences and inherent uncertainties for robust system estimation. Note that the process noise $\omega_{\tau}$ accounts for uncertainties or unpredictable factors affecting the vehicle's motion. The measurement noise $\upsilon_{\tau}$ represents the uncertainty or error associated with sensor measurements. It includes inaccuracies in sensor readings or environmental factors affecting the accuracy of the measurements. The matrices $B$ and $Z$, as well as $\omega_{\tau}$ and $\upsilon_{\tau}$,  collectively form integral components in KF.

Such components handle external factors and uncertainties, laying the groundwork for the two core equations in KF; the process equation (Eq. \ref{eq:kalman_filter-1}) and the measurement equation (Eq. \ref{eq:kalman_filter-2}). The former is used to predict the vehicle's state, whereas the latter is used to integrate new measurements into the filter. Both succinctly captured in Eq. \ref{eq:kalman_filter-2}, these equations represent the predictive and corrective steps essential for the KF's operation.

\begin{subequations}
	\begin{align}
		X_{\tau+1} &= 	A_\tau X_{\tau} + B \omega_{\tau} \label{eq:kalman_filter-1}\\
		Z_{\tau}   &= 	H_\tau X_{\tau} + \upsilon_{\tau} \label{eq:kalman_filter-2}
	\end{align}
	\label{eq:kalman_filter}
\end{subequations}
The KF relies on knowledge of the estimated location of the vehicle from the previous time step to compute accurate predictions. At time $\tau$, the KF models its state using four key parameters. The predicted value based on the previous estimated state is denoted ${\hat{X}^{-}}_{\tau}$. The estimated state obtained from the update step is denoted $\hat{X}_{\tau}$. The predicted covariance matrix based on the previous step's estimated covariance matrix is denoted $P^{'}_{\tau}$.
The estimated covariance matrix obtained from the update step is denoted $P_{\tau}$.

The Kalman prediction involves a two-step process, starting with a prediction step followed by an update step. 
In the prediction step, it initiates with the calculation of the one-step ahead prediction of the state, denoted as ${\hat{X}^{-}}_{\tau}$, as described in Eq. \ref{eq:kalman_filter(predict-1)}. This prediction represents the system's anticipated state at time $\tau$ based on the previously estimated state and system dynamics. Following this, the KF computes the predicted covariance matrix, as shown in Eq. \ref{eq:kalman_filter(predict-2)}. This matrix quantifies the expected uncertainty or error associated with the estimated state at time $\tau$. 

\begin{subequations}
	\begin{align}
		{\hat{X}^{-}}_{\tau} &= 	A_{\tau-1}\hat{X}_{\tau-1} \label{eq:kalman_filter(predict-1)}\\
		P^{'}_{\tau}   &= 	A_{t}P_{\tau-1}A^{T}_{\tau} + Q\label{eq:kalman_filter(predict-2)}
	\end{align}
\end{subequations}

For the update step, the KF first computes the measurement residual, denoted as $\widetilde{y}_\tau$ as shown in Eq. \ref{eq:kalman_filter(update-1)}. The optimal gain $K_\tau$ can then be computed as shown in Eq. \ref{eq:kalman_filter(update-2)}.

\begin{subequations}
	\begin{align}
		\widetilde{y}_\tau &= 	Z_{\tau} - H_\tau{\hat{X}^{-}}_{\tau} \label{eq:kalman_filter(update-1)}\\
		K_\tau   &= 	P^{'}_{\tau}H^{T}_\tau(H_tP^{'}_{\tau}H^{T}_\tau + R)^{-1}\label{eq:kalman_filter(update-2)}
	\end{align}
\end{subequations}

The KF then proceeds by updating the two filter variables, $\hat{X}_{\tau}$ and $P^{'}_{\tau}$, using both $\widetilde{y}_\tau$ and $K_\tau$ as shown in Eq. \ref{eq:kalman_filter(update-3-4)}. The estimated state of $\hat{X}_{\tau}$ is updated with the measurement vector of $Z_{\tau}$ Eq. \ref{eq:kalman_filter(update-3)}. The estimated covariance matrix $P_{\tau}$ is then updated accordingly Eq. \ref{eq:kalman_filter(update-4)}.

\begin{subequations}
	\begin{align}
		\hat{X}_{\tau} &= 	\hat{X}^{-}_{\tau} + K_\tau\widetilde{y}_\tau \label{eq:kalman_filter(update-3)}\\
		P_{\tau}   &= 	(I - K_\tau H_\tau)P^{'}_{\tau}\label{eq:kalman_filter(update-4)}
	\end{align}
	\label{eq:kalman_filter(update-3-4)}
\end{subequations}

Upon completion of both the prediction and update steps,
the KF computes the optimal estimate value $\hat{X}^{-}_{\tau}$ for $X_\tau$. This computation takes into account both the measurement uncertainty and process noise.
This optimal estimate value $\hat{X}^{-}_{\tau}$ is then utilized in the subsequent iterative prediction step, instead of using the real observed value. 

After estimating the next location of the vehicle, a crucial step involves comparing this estimated location with the current heading direction of the vehicle. This comparison serves as the basis for determining the subsequent heading direction, indicating whether the vehicle is expected to turn left or right, as illustrated in Figure \ref{fig:ROI estimation utilizing Kalman filter}. This directional insight plays a pivotal role in estimating the ROI for the vehicle. The ROI is designed as a triangular region aligned with the anticipated turning direction, providing essential guidance for the vehicle during its navigation process.

%% file: 2d.problem_formulation.tex


\subsection{Quality Utility Evaluation Module}~
Upon estimating the ROI for each vehicle, these estimations are utilized to calculate the utility function, which measures the VOI. In Q-CPTO, a metric for traffic perception denoted as $a_{ij}$ evaluates users' detected traffic awareness based on the non-redundant perception area of user $u_i$. This area quantifies the percentage of space covered by both the sensors of user $u_i$ and the aggregated results of its collaboration with worker $w_j$, as depicted in Figure \ref{fig:step-1}.
The computation of $a_{ij}$ involves projecting the triangular coverage area of a vehicle's sensor FOV onto a projection matrix, denoted $\mathbb{M}$. Subsequently, triangular rasterization is applied, assigning a value of 1 to each cell in the matrix intersecting the triangular FOV and 0 to the remaining cells. Notably, worker $w_j$ engages in collaborative efforts with multiple users, acquiring their diverse perceptions for aggregation. The aggregated output is then transmitted to the associated users, thereby expanding their overall perception capabilities.

Furthermore, in our approach, we elevate the detected traffic awareness of individual users by incorporating the concept of shared interest, denoted $q_{ik}$, among user pairs $u_i$ and $u_k$ (where $i \in U, k \in U, i \neq k$). To compute $q_{ik}$, user $u_i$ employs a triangular ROI projection, denoted $\mathbb{T}$, based on the predicted ROI by the Vehicle Mobility Prediction module, as illustrated in Figure \ref{fig:step-2}. Subsequently, a multiplication operation is performed between the perceptions of all users in the system and $\mathbb{T}$ as depicted in Figure \ref{fig:step-2}. This computation results in the determination of $q_{ik}$. The vehicles that contribute to the highest utility, specifically those with the most substantial masked non-redundant perception areas, are selected to offload to the same worker (i.e., cooperate together), as illustrated in Figure \ref{fig:step-3}. Our primary objective is to search for the optimal groups of cooperating vehicles, known as platoons, that maximize their respective perceptions. In essence, this study endeavors to identify the vehicle platoons that yield the most extensive masked non-redundant perception areas, thereby enhancing traffic awareness for participating users.
 
\subsection{Problem Formulation}
Q-CPTO is designed to enhance the perceived traffic awareness of the users in the system by maximizing the shared interest between users. For a given binary placement decision variable $x_{ij}$, represented as 1 when $u_i$ offloads to $w_j$ and 0 otherwise (as defined in Eq. \ref{eq:vars-x}), Q-CPTO aims to maximize the placement.
The shared interest metric $q_{ik}$ quantifies the quality relationship between users $u_i$ and $u_k$ (as detailed in Eq. \ref{eq:vars-q}). 

\begin{subequations}
	\begin{align}
		X &= 	\{x_{ij} | i \in U, j\in W \} \label{eq:vars-x}\\
		Q &= 	\{q_{ik} | i \in U, k\in U, i\neq k \} \label{eq:vars-q}
	\end{align}
\end{subequations}

Q-CPTO's objective is to maximize the weighted-sum shared interest placement of users within the system. This objective is expressed in Eq. \ref{eq:Q-CPTO}.
This approach ultimately enhances the users' perception of traffic conditions.

\begin{smaller}
	\begin{subequations}
		\begin{align}
			\max_{x} 	  \quad & \sum_{j=1}^{m}  \left( \sum_{i=1}^{n-1}\sum_{k=i+1}^{n}     (x_{ij}x_{kj})q_{ik} \right)	\label{opti:latency}\\
			\textrm{s.t.} \quad 
			&\left(  (\sum_{i=1}^{n}{x_{ij}})\right)   + \left((1 - x_{ij}) \times 2 (2)\right)  \ge 2	&\tiny{ \forall  j \in W \quad \forall  i \in U}
			\label{opti:perception_constraint}\\	
			& \sum_{i=0}^{n}{x_{ij}t_{ij}\le \kappa_i}
			& \forall  j \in W
			\label{q-cpto:deadline}\\
			& \sum_{j=0}^{m}{x_{ij}\le 1} 	
			&
			\forall i \in U 	\label{q-cpto:user_served_by_1_worker}\\
			& x_{ij} \in \{0,1\}  
			&\quad \forall i \in U \quad \forall j \in W \label{q-cpto:placement}
		\end{align}
		\label{eq:Q-CPTO}
	\end{subequations}
\end{smaller}

This weighted sum allows Q-CPTO to cluster users based on their mutual potential to enhance their individual's traffic perceptions. Vehicles within the same cluster offload their perceptions to the same worker, enabling the perception aggregation process. This selective offloading strategy optimizes traffic perception while conserving network and computation resources.

Q-CPTO operates at the SD-CSO; it adapts to the dynamic network and traffic conditions, which evolve over discrete time slots $\mathcal{T}$.
To identify the optimal solution under the varying system status, Q-CPTO focuses on a short time point $\tau \in \mathcal{T}$, where traffic and network conditions are considered relatively constant and serve as a benchmark.
During each $\tau \in \mathcal{T}$, SD-CSC adds users to a virtual queue $\mathbb{U}$ when it estimates, using the KF, that a user will be turning either left or right. Subsequently, Q-CPTO evaluates which users in the $\mathbb{U}$ queue require offloading.
In Q-CPTO, the first constraint constraint \ref{opti:perception_constraint} imposes a lower bound on the number of collaborating users; it mandates that a minimum of two users collaborate for any valid task offloading.
Constraint \ref{q-cpto:deadline} guarantees that the response delay of the perception task of user $u_i$ on worker $w_j$ does not exceed a certain threshold (i.e., deadline) $\kappa_i$.
Constraint \ref{q-cpto:user_served_by_1_worker} ensures that each user is served by at most one worker.
Constraint \ref{q-cpto:placement} is the integrity constraint associated with the binary decision variable $x_{ij}$.

The formulation of Q-CPTO can be modeled as a quadratic multiple knapsack problem (QMKP) \cite{QMKP-survey}. The QMKP is a combination of two NP-hard combinatorial optimization problems, the Quadratic Knapsack Problem QKP \cite{QKP} and the Multiple Knapsack Problem MKP \cite{multknaps}.
QMKP has a set of knapsacks $M$ with fixed capacities $C$ and a set of items $N$,
each characterized by a profit and a weight.
 QMKP aims to assign items to knapsacks, maximizing the joint profit $p_{ij}$ of items $\{i \in N, j \in N, i \neq j\}$ while adhering to the knapsacks' capacity constraints. In modeling Q-CPTO as QMKP, the shared interests among users, denoted $q_{ik}$, are interpreted as the joint profit within QMKP. 
The users weights are modeled as their latency $t_{ij}$, and the worker's capacity constraint is established by modifying constraint \ref{q-cpto:deadline} to Eq. \ref{q-cpto:worker-capacity}.

\begin{equation}
	w_{c_j} =   \frac{\kappa_i C_j}{\overline{l_i}} 
	\label{q-cpto:worker-capacity}
\end{equation}

The worker capacity, denoted as $w_{c_j}$, is determined based on the current computation intensity of worker $w_j$, represented as $C_j$, alongside the task deadline $\kappa_i$ and the average computation intensity of perception tasks conducted by users in the problem, denoted ${\{ \overline{l_i} | i \in U \}}$. This equation essentially establishes an upper bound on the number of users eligible to offload their tasks to worker $w_j$.

\begin{algorithm}
	\hspace*{\algorithmicindent} \textbf{Input}: Quality Matrix (Q), Workers Capacities ($w_c$)\\
	\hspace*{\algorithmicindent} \textbf{Output}: WorkerAssignments
	\caption{:Q-CPTO-H}
	\label{alg:Q-CPTO-H}
	\begin{algorithmic}[1]
		\Procedure{greedyBinPacking}{$C$}
		\State Set $J^{\prime}=N$ and $\bar{c}=c:=\left(c_{1}, \ldots, c_{m}\right)$;
		\State Sort the indices of $N$ in decreasing order of their densities, i.e., $d_{1} \geq d_{2} \geq \ldots \geq d_{n}$;
		\State Sort all workers in decreasing order of their capacities, i.e., $\bar{c}_{1} \geq \bar{c}_{2}, \ldots, \geq \bar{c}_{m}$;
		\State Set $\ell=0 ; / /(\ell+1)$ is the first user to pack into the first worker
		\Repeat
		\State Increment $(\ell)$;
		\State if $\left(\exists \rho \mid \bar{c}_{\rho} \geq w_{\ell}\right) / /$ according to the ranking used on the vector $\bar{c}$ ) then
		\State $\quad$ Set $x_{\ell \rho}^{\star}=1$ and $\forall k \in M \backslash\{\rho\}, x_{\ell k}^{\star}=0$;
		\State $\quad$ Set $\bar{c}_{\rho}=\bar{c}_{\rho}-w_{\ell}$;
		\State else
		\State $\quad$ $\operatorname{Set} x_{\ell k}^{\star}=0, \forall k \in M$;
		\State end if
		\State Set $J^{\prime}=J^{\prime} \backslash\{\ell\}$;
		\Until $\left(J^{\prime}=\emptyset\right)$;
		\EndProcedure
		
		\State $S$ $\gets greedyBinPacking(W_c)$;
		\State Set $S^{\star}=S$ and generate $\alpha$ in the interval ] $0 \%, 100 \%[$
		\Repeat
		\State Let $S_{1}$ be the set having $\left(\left|S^{(1)}\right| \backslash \alpha\right)$ of variables fixed to 1 (a random copy of users of $S$ );
		\State Let $S_{2}=N \backslash S_{1}$ and $S_{2}^{\prime}$ $\gets greedyBinPacking()$ as a complementary solution when applied to the set of free workers $S_{2}$.
		\State Set $S^{\prime}:=S_{1} \cup S_{2}^{\prime}$ as the new solution built.
		\State Update $S^{\star}$ with $S^{\prime}$ if $z\left(S^{\prime}\right)>z\left(S^{\star}\right)$;
		\State Set $S=S^{\star}$ if $\operatorname{rand}() / 2=0$;
		\Until (satisfying the stopping condition)
		\State return $S^{\star}$

	\end{algorithmic}
\end{algorithm}

Due to the high complexity of Q-CPTO, a heuristic approach is employed to efficiently solve the problem. This approach aims to find a near-optimal assignment in a time-efficient manner.

\subsection{Q-CPTO-Heuristic (Q-CPTO-H)}
To address the aforementioned problem, we leverage a reactive local search designed by Hifi and Michrafy \cite{QMKP-FCS}. This approach involves degrading the solution to escape local optima and diversify the search process. A memory list based on a hashing function is used to prevent repeated configurations and improve solution quality.
The procedure follows a Fix and Complete approach (FCS), which combines two main strategies: a fixing strategy and a greedy assignment strategy. The fixing strategy leverages a greedy bin packing technique for constructing an initial solution, completing the current solution, or rectifying an unfeasible solution. The greedy assignment strategy uses a hashing function to mimic the tabu search strategy.
Combining these techniques, Q-CPTO-H effectively navigates the solution space, and augments solution quality for intricate optimization challenges.

The Q-CPTO-H method begins by using a greedy bin packing approach to construct an initial feasible solution, denoted $S=\left(s_{1}, \ldots, s_{n}\right)$, for the assignment problem. Specifically, it assigns users to workers based on their computation capacity until this capacity is fully utilized.
Next, two sets, $N_{1}(S)$ and $N_{0}(S)$ are created, representing subsets of users with values of 1 and 0, respectively.
Additionally, Q-CPTO-H introduces two helper functions. The first is the contribution of user $\{u_i, i \in N\}$  to the objective function for the feasible solution $S$, denoted $c_{i}(S)$ shown in Eq. \ref{q-cpto-helper:contribution}, which is the received augmented perception of $u_i$ when offloading to worker $w_j$.
\begin{equation}
		c_{i}(S)=\sum_{\substack{j \in M, i \neq j}} q_{i j} \label{q-cpto-helper:contribution}
\end{equation}
 The second is the density of user assignment $\{u_i, i \in N\}$, according to both the feasible solution $S$ and the worker $\{w_j, j \in M\}$, denoted $d_{i}(S, w_j)$ shown in Eq. \ref{q-cpto-helper:density}, which is determined by dividing both the contribution of $u_i$ to the objective function, on the weight of assigning $u_i$ to the worker $w_j$ (i.e, delay of offloading to worker $t_{ij}$).
 \begin{equation}
 	d_{i}(S, w_j)=\frac{c_{i}(S)}{t_{ij}}\label{q-cpto-helper:density}
 \end{equation}
In summary, Q-CPTO-H constructs a feasible solution using a greedy approach, creates subsets of user-worker associations based on their values, and utilizes helper functions to compute users' contributions and the density of each user assignment with respect to the objective function and worker capacity.

First, the greedy bin packing approach is outlined in Algorithm \ref{alg:Q-CPTO-H}. This function initiates by setting all user-worker associations to free (line 2), and defining the vector $\bar{c}$ as the residual capacity equivalent to $c$. Subsequently, it sorts all users in descending order based on their densities (line 3), as determined by Eq. \ref{q-cpto-helper:density}. In a similar fashion, the contribution capacities are sorted in a descending order, provided they have distinct values (line 4), employing Eq. \ref{q-cpto-helper:contribution}. This approach ensures an optimized packing process by granting precedence to users with higher densities and larger contribution capacities.

Subsequently, the core assignment procedure initiates a loop (lines 6-15). In this loop, an unassigned user, denoted $u_\ell$, with the highest density $d_{\ell}$ is chosen and added to the partial solution. Concretely, the $u_\ell$-th user is assigned to an open worker, denoted $\rho$. This assignment occurs if the weight of the user $w_{\ell}$ (i.e., delay) is less than or equal to the residual capacity $\bar{c}_{\rho}$ of the worker. If this condition is not met, the user is either allocated to a newly assigned worker or its value is fixed at zero for all workers. This iterative process  persists until all users have been evaluated, at which point it returns the best found solution denoted $x^{\star}$.

After reaching an initial solution, Q-CPTO-H builds upon the found solution through adopting a fixing strategy. This strategy involves the formation of two subsets, $S_{1}$ and $S_{2}$, whose union encompasses all $n$ users and their intersection is an empty set (i.e., $\left|S_{1} \cup S_{2}\right|=n$ and $S_{1} \cap S_{2}=\emptyset$). 
Specifically, $S_{1}$ represents a subset that contains already fixed variables, while $S_{2}$ is a subset of unfixed or free variables.
These subsets are generated by assuming a feasible solution is denoted by $S$. A percentage $\alpha$ is then generated within the interval of $] 0 \%, 100 \%[$,
facilitating the subsequent construction of $S_{1}$ and $S_{2}$.
Initially, $S_{1}$ is created as a copy of a feasible solution $S$, from which $\alpha \times |S^{(1)}|$ variables are randomly dropped, where $|S^{(1)}|$ denotes the number of components of $S$ fixed to 1. This results in $S \backslash \alpha$ variables being fixed to 1 in $S_{1}$.
Subsequently, $S_{2}$ is constructed by setting it equal to $N \backslash S_{1}$. The greedy packing procedure is then applied to $S_{2}$, resulting in a new solution denoted $S_{2}^{\prime}$ containing user associations fixed to 1 or 0.
Finally, a new solution $S^{\prime}$ is formed by the union of $S_{1}$ and $S_{2}^{\prime}$. This systematic approach allows for efficient handling and assigning variables within the subsets.

Algorithm \ref{alg:Q-CPTO-H} incorporates the previously mentioned procedure (line 17) into a descent stage.
This stage begins with removing $\alpha$ user associations from the current solution (lines 20--22), which forms $S_{1}$. A complementary solution, $S_{2}^{\prime}$, is then generated using the set $S_{2}$, which contains both the free variables and a local solution $S^{\prime}$. This local solution is a combination of $S_{1}$ and $S^{\prime}$. 
Both the local solution and the best solution found so far are updated as necessary. 
This iterative procedure continues until the final stopping condition is satisfied.

The systemic approach adopts a descent method and involves a degree of randomness in the selection of the incumbent solution necessitating an alternative choice between the best solution reached, denoted $S^{\star}$, and the current solution, $S$. Specifically, a random function (line 24) is employed, offering a 50\% probability of choosing either $S^{\star}$ or $S$ for perturbation. This element of randomness allows for the experimentation of different solutions, thereby leading to improved optimization results.

%% file: 2e.experiments.tex
\section{Performance Evaluation}
\label{sec:exps}

In this section, we present a comprehensive evaluation of the Q-CPTO-H. Our focus is on assessing its performance in optimizing cooperative perception and aggregation latency. To gain a deeper understanding, we compare Q-CPTO-H against a representative set of state-of-the-art optimization-based task allocation schemes, specifically those emphasizing latency optimization but overlooking the cooperative perception aspect.

To demonstrate the effect of considering
cooperative perception in Q-CPTO-H, we evaluate its performance to three offloading strategies: (1) exhaustive optimal search (Q-CPTO), which optimally solves the offloading problem; (2) greedy offloading scheme (GO) \cite{greedy-baseline}, which offloads all perception tasks up to the maximum worker capacity, imposing an upper limit on system latency; and (3) CPTO \cite{ICC2023-CPTO}, which is designed to minimize response latency while maximizing the number of offloading users, employing a uniform offloading approach \cite{uniform-offloading}. Through this comparative analysis, our goal is to derive valuable insights into the effectiveness of Q-CPTO-H in optimizing traffic perception and response latency.

The experimental results are organized into four key segments:
1) the proposed schemes are compared with several
baselines while varying the computational capacity of the workers; 2) the impact of the computation latency
threshold (deadline) on the evaluated metrics; 3)
the implications of an increased
number of users on system performance; 4)
and finally we study the temporal performance of offloading decision latency for
both Q-CPTO and Q-CPTO-H, considering a
varying number of system users. This comprehensive analysis
provides valuable insights into the system’s adaptability and
efficiency under various conditions.

We use the following performance metrics: First, the average detected awareness metric assesses the quality of the CP service by measuring the percentage of the ROI coverage achieved through aggregating perceptions from collaborating vehicles and the vehicle's own perception. Second, we analyze the average response latency (delay), which begins when a perception frame request is dispatched to a worker and ends when a response is received. Third, we consider the average number of collaborating vehicles associated with workers, which provides insights into the level of collaboration in the system \cite{ICC2023-CPTO}\cite{PLTO}\cite{7-data-redundancy}. Finally, we examine the traffic perception satisfaction ratio, denoted as $\varphi$, which represents the percentage of requests resulting in a perception that meets or exceeds the required ROI coverage.

\begin{figure*}[ht!]
	\subfloat[Avg Detected Awarness \label{fig:exp-cpu-awarness}]{%
		\includegraphics[ width=0.25\textwidth]{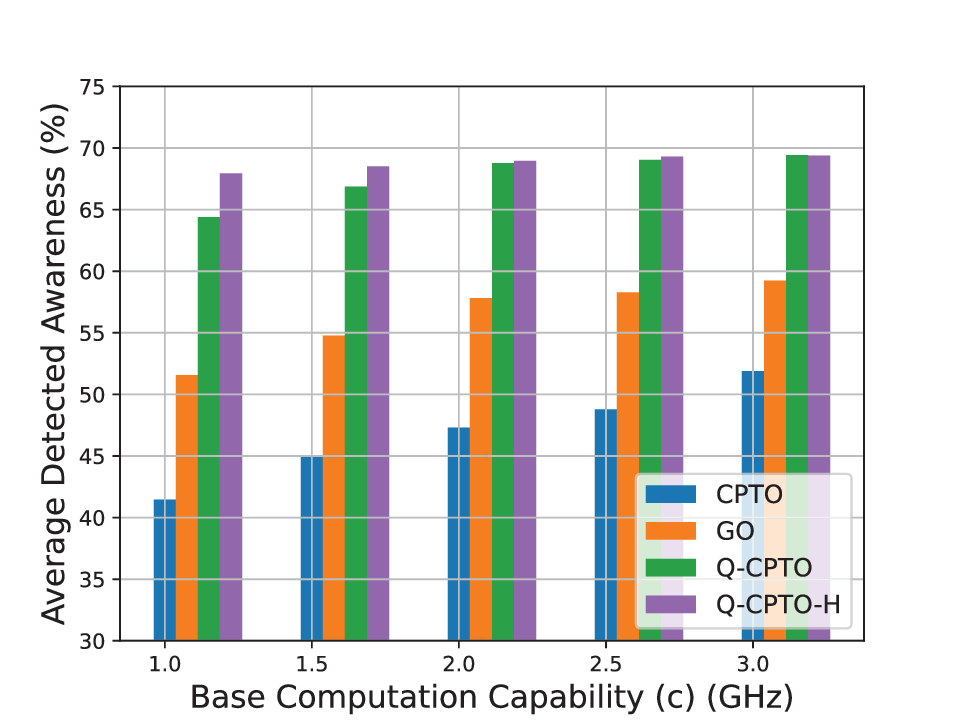}}
	\subfloat[Avg Perception Intensity \label{fig:exp-cpu-pi}]{%
		\includegraphics[ width=0.25\textwidth]{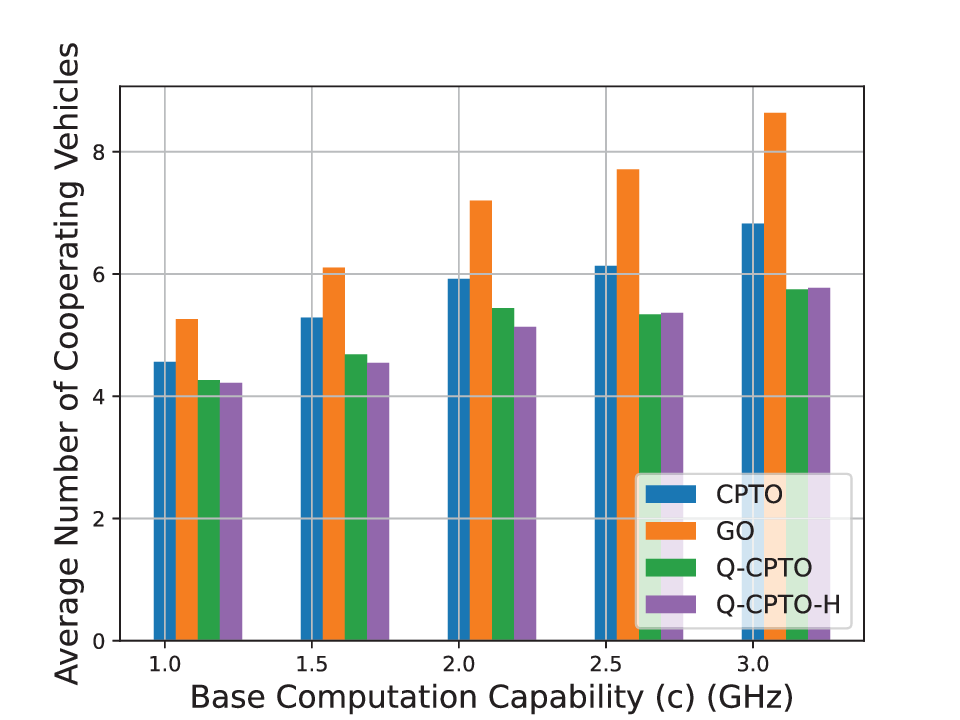}}
	\subfloat[Avg Delay \label{fig:exp-cpu-delay}]{%
		\includegraphics[ width=0.25\textwidth]{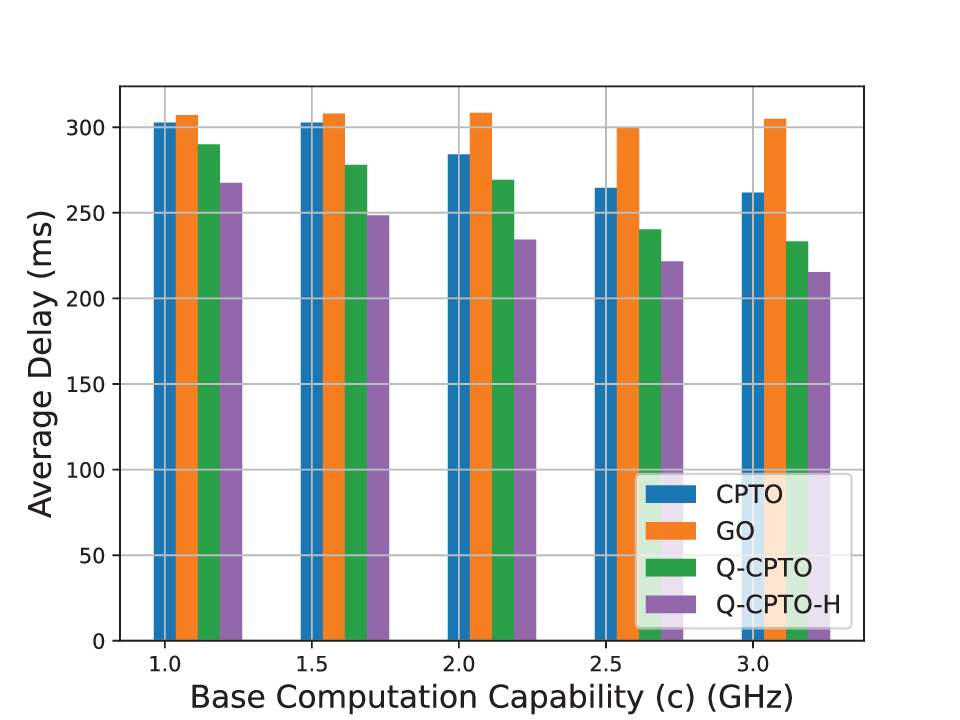}}
	\subfloat[Satisfaction Ratio \label{fig:exp-sat-cpu}]{%
		\includegraphics[ width=0.25\textwidth]{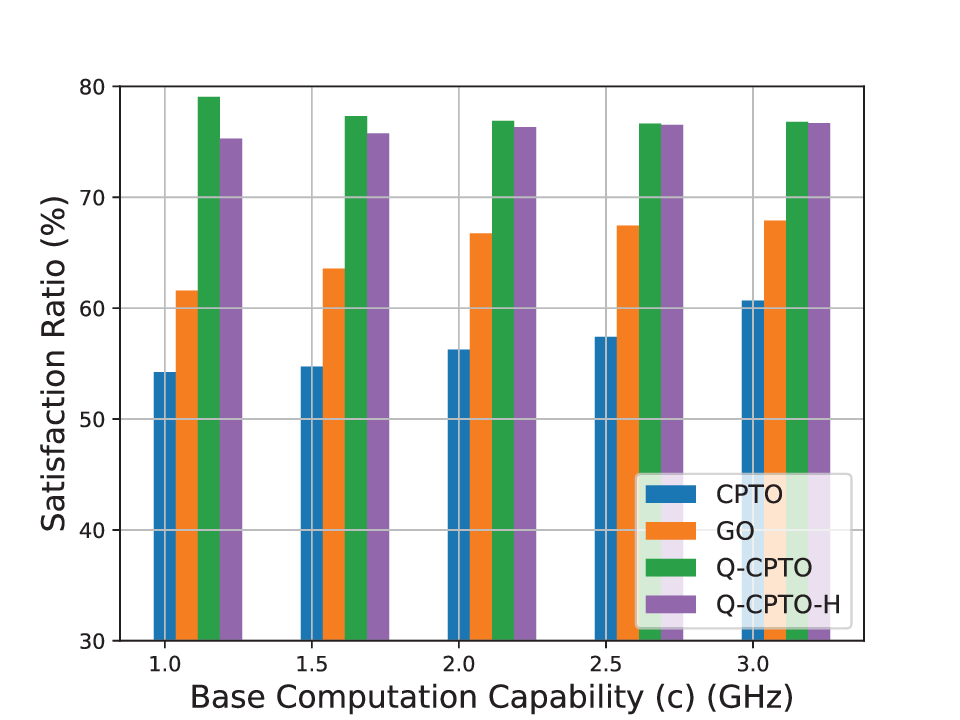}}\\
	\caption{Impact of the schemes  over varying computation capability (deadline=0.4, users=40).}
	\label{fig:exp-cpu}
\end{figure*}

\subsection{Simulation Setup}

We have implemented GO, CPTO, Q-CPTO, and Q-CPTO-H schemes using Python \footnote{The source code will be made available upon acceptance of this paper.} 
and integrated the simulator with IBM CPLEX
optimization solver \cite{IBMcplex} to solve the optimization problems.

Vehicular mobility traces are created using the Simulation of Urban MObility (SUMO) traffic
simulator \cite{SUMO}. The network is designed with a dimensions of 200x200 meters, and features a two-lane multi-directional turn with two lanes in each direction. Vehicles simulations are conducted at a maximum speed of 40 Km/h over a total runtime of 60 seconds.  
We place eight RSUs uniformly along the various lanes, each with a communication range assumed to be 200 m \cite{FogFollowingMe}. 
In our simulation setup, each RSU is linked to a worker for offloading tasks. The computational resources of these workers are randomly chosen from the set \{c + $1\delta$, c + $2\delta$, c + $3\delta$\}, where c is set at 2 GHz, and $\delta$ equals 1 GHz unless otherwise specified. Furthermore, to simulate a range of user transmission capabilities, the uplink data rate of the vehicles spans from 15 to 18 Mbps, including both low and high transmission speeds  \cite{DSRC-1}.

In our perception task simulations, vehicles are equipped with state-of-the-art LiDAR sensors, producing approximately 100,000 points per frame, equivalent to around 4 MB of data per frame \cite{F-Cooper}. The transmission of such substantial data volumes poses challenges to wireless networks. To address this, we leverage recent advancements in 3D point cloud object detection, employing Convolutional Neural Networks (CNNs) to generate and transmit only the essential features. These features, as indicated by \cite{F-Cooper}, can be efficiently compressed to 200 Kb, taking only a few milliseconds, while maintaining the accuracy of 3D object detection algorithms.

To further enhance perception capabilities, we incorporate advanced point cloud fusion techniques detailed in \cite{F-Cooper}, extending the detection range to 50 m \cite{CoFF}. To maintain conservative estimates, we prudently set the detection range for the employed fusion technique to 20 m. Additionally, the transmitted data ($\lambda_i$) is set to 200 Kb. The computational intensity ($l_i$) is established at $2 \times 10^8$ cycles \cite{Case_Cooperative_Perception_globecomm2020}.

The ROI as an equilateral triangle with a height of 10 m in the direction of the vehicle's estimated turn, we maintain a perception latency threshold of 0.4 seconds \cite{Case_Cooperative_Perception_globecomm2020}. To ensure robust ROI coverage, the lower bound denoted as $\varphi$ is set at 35\%, surpassing the perception achieved by vehicles without offloading. 
The simulation period is set to 60 seconds, and the optimization problem is addressed periodically by the SD-CSO every 1 second. The number of users engaged in the offloading optimization process varies in each run of the decision-making schemes. Specifically, only the users who are estimated to take a left or a right turn are considered by SD-CSO in the offloading schemes.

\begin{figure*}[ht!]
	\subfloat[Avg Detected Awarness\label{fig:exp-deadline-awarness}]{%
		\includegraphics[ width=0.25\textwidth]{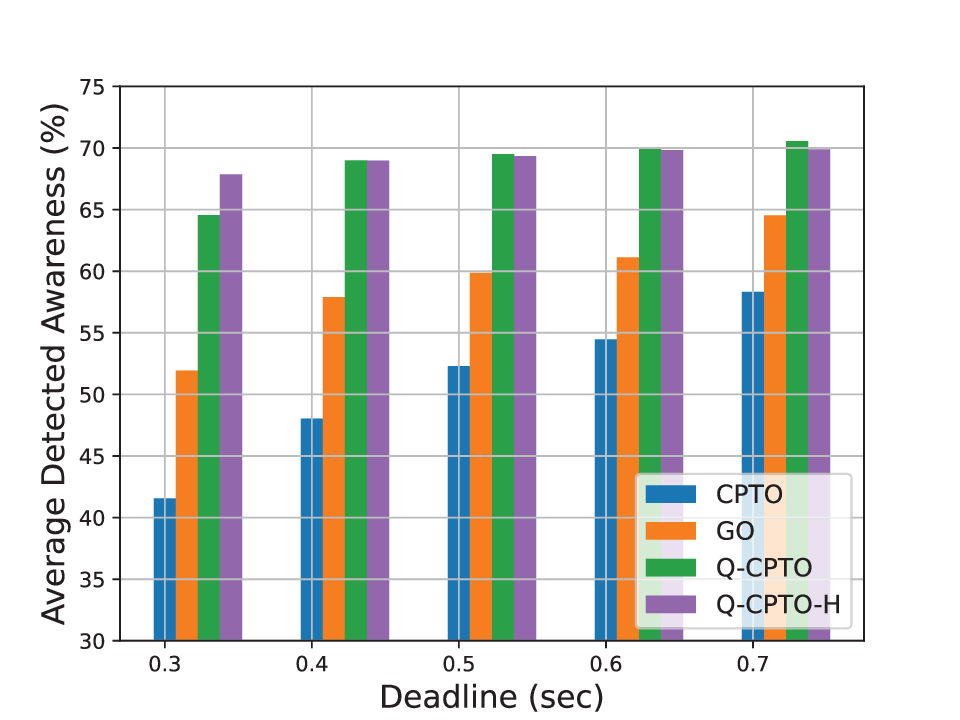}}
	\subfloat[Avg Perception Intensity \label{fig:exp-deadline-pi}]{%
		\includegraphics[ width=0.25\textwidth]{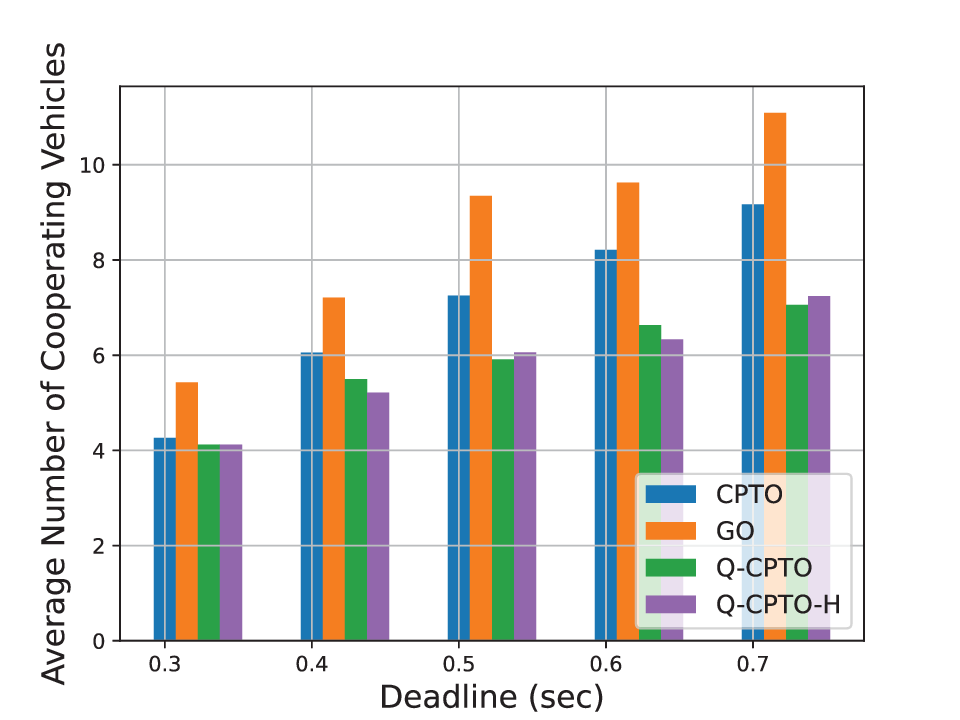}}
	\subfloat[Avg Delay \label{fig:exp-deadline-delay}]{%
		\includegraphics[ width=0.25\textwidth]{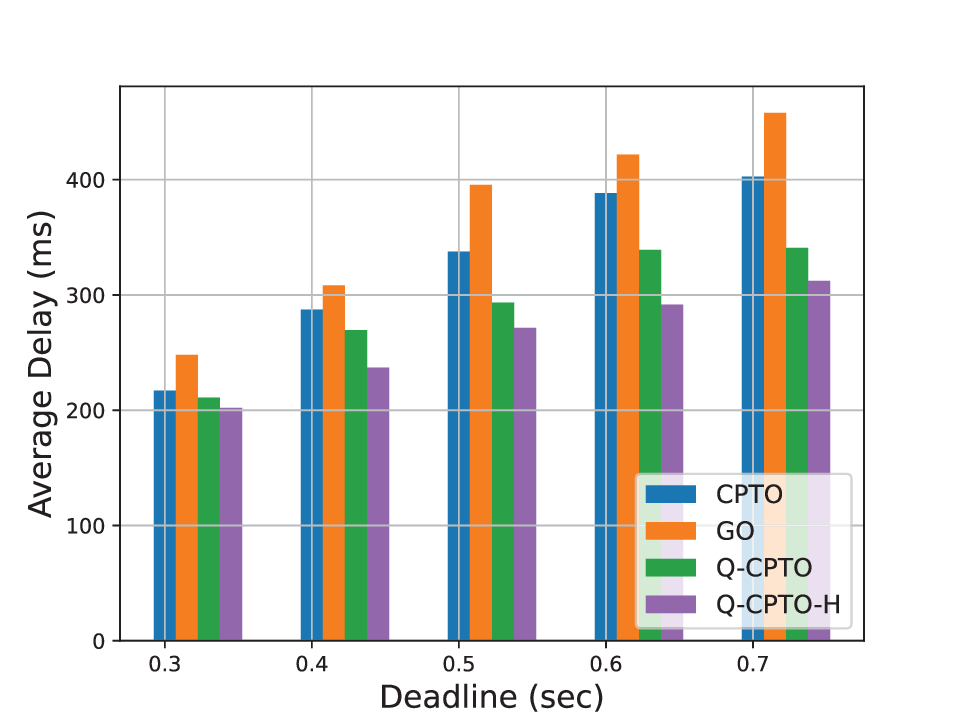}}	
	\subfloat[Satisfaction Ratio \label{fig:exp-deadline-sat}]{%
		\includegraphics[ width=0.25\textwidth]{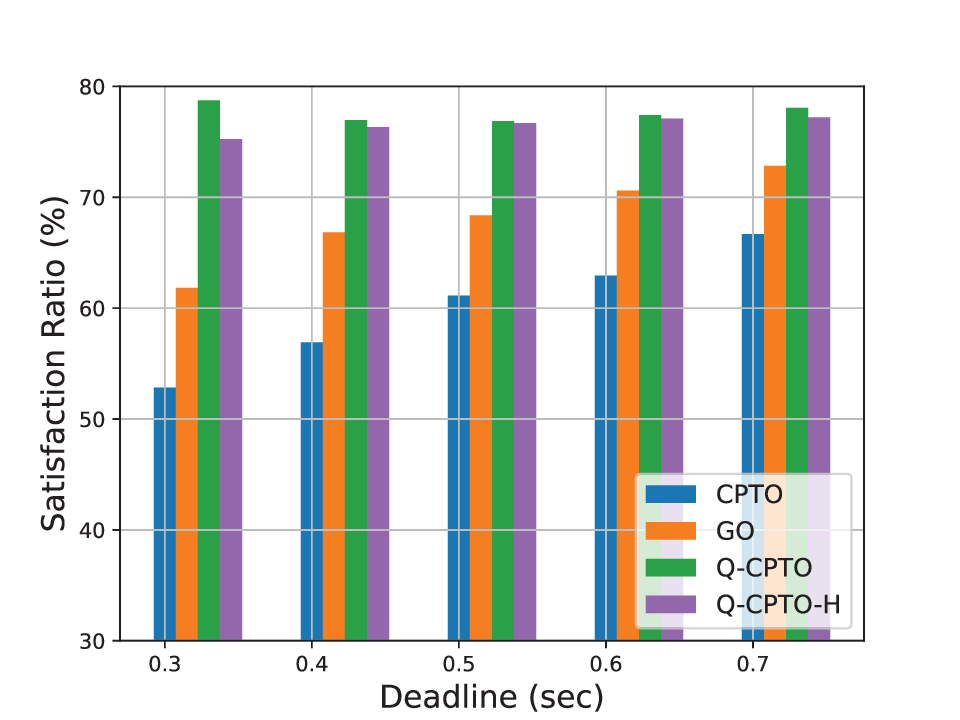}}\\
	\caption{Impact of the schemes  over varying deadlines (c=2, users=40).}
	\label{fig:exp-deadline}
\end{figure*}

\subsection{Results and Analysis}
In this section, we present a detailed analysis of the performance of the four distinct offloading schemes GO, CPTO, Q-CPTO, and Q-CPTO-H across a range of experiments. The experiments explore the impact of varying computational capacity, perception task deadlines, and the number of connected users on the cooperative perception system. By evaluating key metrics such as average detected awareness, perception intensity, response delay, and satisfaction ratios, we aim to provide a comprehensive understanding of each offloading scheme. Moreover, we provide a time performance analysis of the offloading decision-making process, providing valuable insights into the scalability and efficiency of Q-CPTO-H. Simulation results are presented at a confidence level of 95\%.

\subsubsection{Impact of varying computational capacity}
In this experiment, depicted in Figure \ref{fig:exp-cpu}, the base computational capacity of the workers ($c$) is varied to examine the impact of increasing system computational resources across various schemes. We vary $c$ from 1.0 to 3.0 GHz while maintaining a constant user count of 40 and holding all other parameters fixed. 

The results, as shown in both Figure \ref{fig:exp-cpu-awarness} and Figure \ref{fig:exp-sat-cpu}, demonstrate that increasing the available computational resources ($c$) in the system leads to a higher average detected awareness for all offloading schemes, including GO, CPTO, Q-CPTO, and Q-CPTO-H. This effect stems from the fact that with enhanced computational resources in workers, a greater number of users can offload to the same worker within the allotted latency threshold (deadline), as illustrated in Figure \ref{fig:exp-cpu-pi}, resulting in the fusion of more perceptions and an expanded coverage area of the ROI. 
Moreover, with the increase in resource availability, the response delay of the fusion operation tends to decrease, as evident in Figure \ref{fig:exp-cpu-delay}.

Our analysis reveals that both Q-CPTO and Q-CPTO-H consistently outperform GO in terms of average detected awareness, as demonstrated in Figure \ref{fig:exp-cpu-awarness}, while employing fewer collaborating users, as shown in Figure \ref{fig:exp-cpu-pi}. Specifically, Q-CPTO and Q-CPTO-H achieve an average detected awareness higher than that of GO by up to 20.3\% and 22.4\%, respectively, while employing 26.1\% and 27.5\% fewer collaborating users. This advantage arises because these quality-driven schemes prioritize offloading users based on their potential contribution to awareness, whereas GO offloads users to workers without considering their VOI. Consequently, this results in a reduction of delay by 14.2\% and 22.3\% compared to GO, primarily due to a decrease in the number of users sharing a given worker, as depicted in Figure \ref{fig:exp-cpu-delay}. Moreover, when compared to CPTO, which aims to minimize response delay while maximizing the number of offloaded requests, Q-CPTO and Q-CPTO-H achieve significantly higher traffic awareness, surpassing CPTO by 44.9\% and 47.5\%, respectively. These results reflect the substantial impact of selective offloading strategies on cooperative perception, while resulting in a 7.5\% and 16.2\% reduction in response delay compared to CPTO.

Furthermore, our analysis underscores the superior satisfaction ratios achieved by the quality-driven schemes when compared to the studied baselines, as evident in Figure \ref{fig:exp-sat-cpu}. Specifically, Q-CPTO and Q-CPTO-H achieve higher awareness satisfaction ratios, surpassing GO by 18.3\% and 16.4\%, respectively, and outperforming CPTO by 36.7\% and 34.5\%, respectively. 

Notably, Q-CPTO-H demonstrates performance results similar to those of Q-CPTO, with small performance gaps of 1.7\%, 2.1\%, 9.3\%, and 1.5\% in terms of average detected awareness, number of collaborating users, response delay, and satisfaction ratio, respectively. These findings collectively affirm the effectiveness of quality-aware offloading strategies in enhancing CP, ultimately leading to higher satisfaction among users in the system.

\subsubsection{Impact of varying the perception task deadline}~\\
\begin{figure*}[ht!]
	\subfloat[Avg Detected Awarness\label{fig:exp-users-awarness}]{%
		\includegraphics[ width=0.25\textwidth]{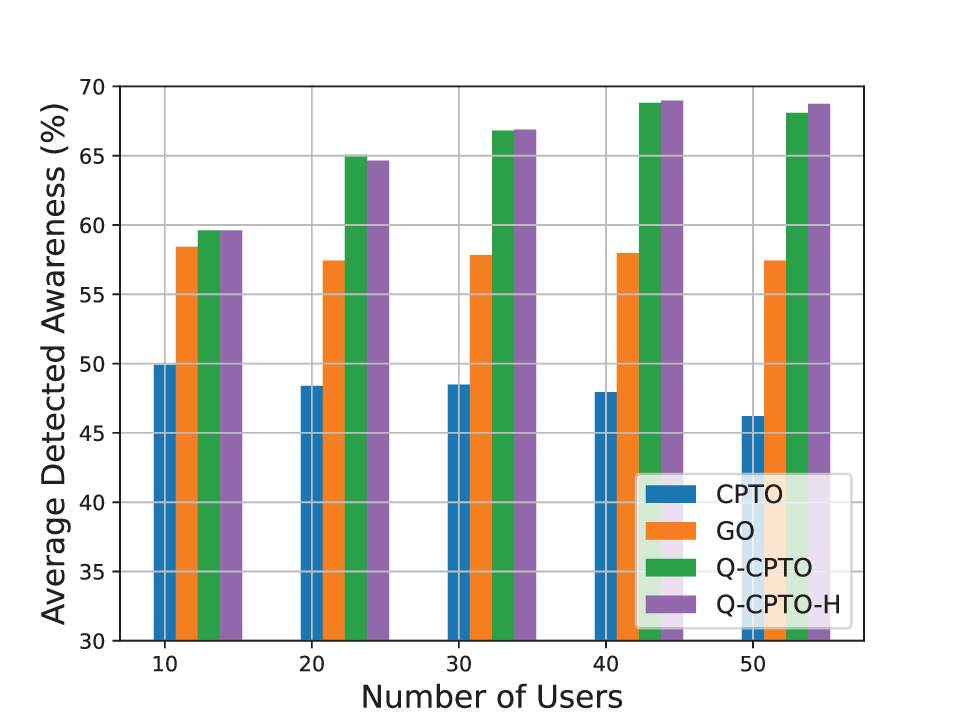}}
	\subfloat[Avg Perception Intensity \label{fig:exp-users-pi}]{%
		\includegraphics[ width=0.25\textwidth]{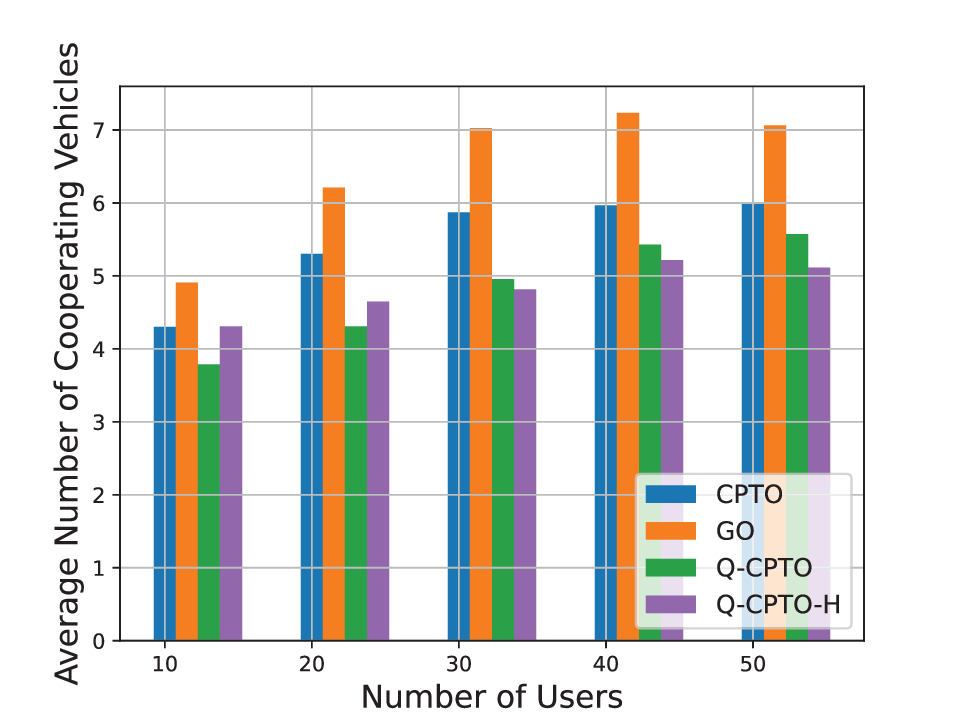}}
	\subfloat[Avg Delay \label{fig:exp-users-delay}]{%
		\includegraphics[ width=0.25\textwidth]{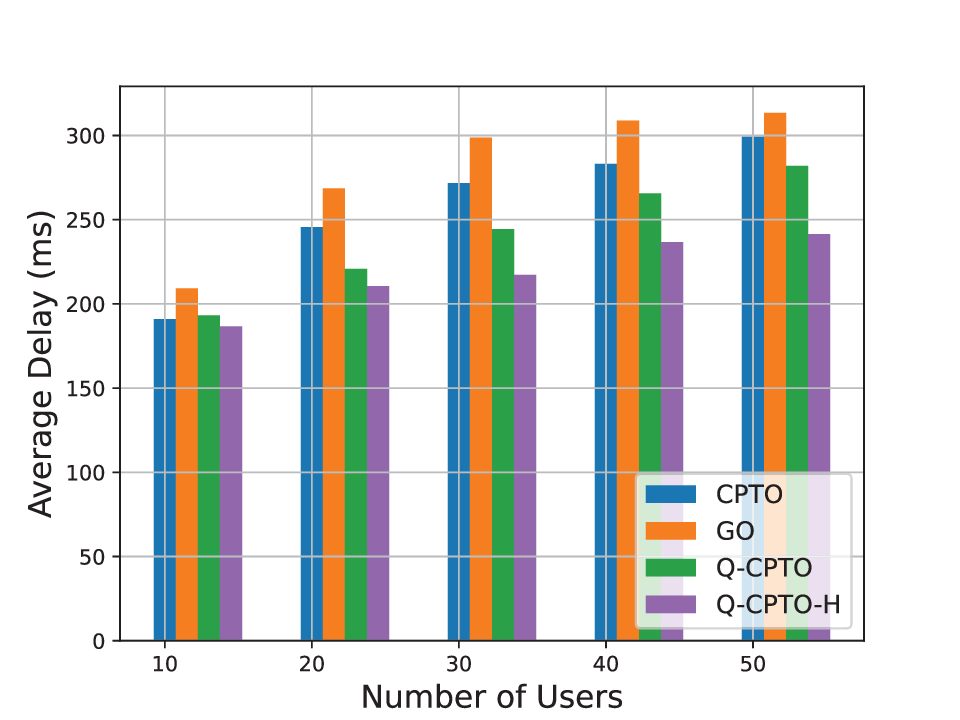}}
	\subfloat[Satisfaction Ratio \label{fig:exp-users-sat}]{%
		\includegraphics[ width=0.25\textwidth]{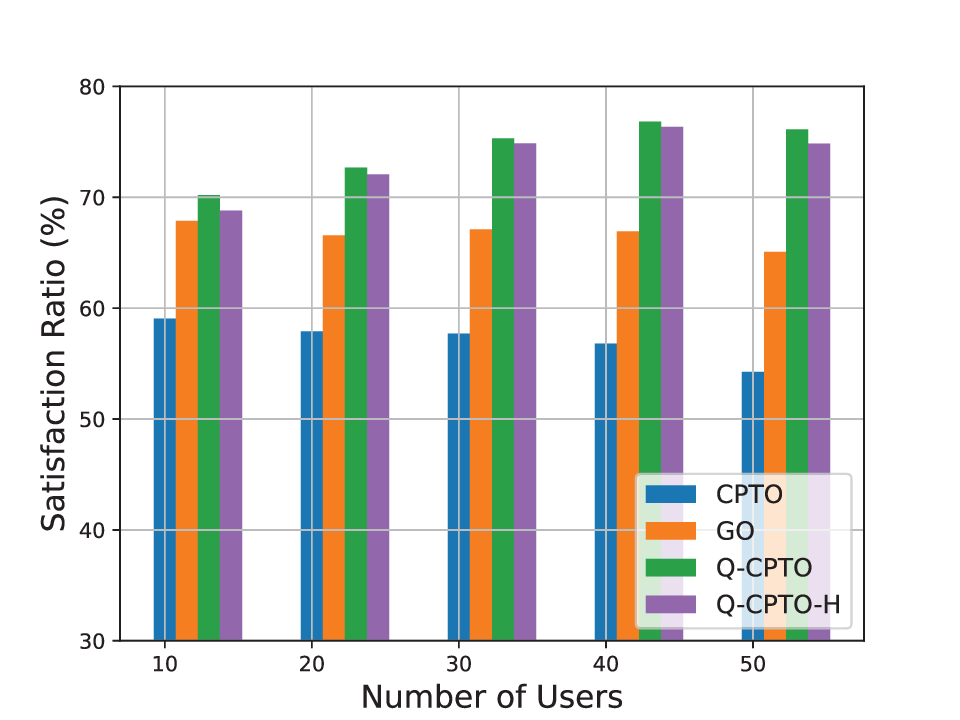}}\\
	\caption{Impact of the schemes  over varying users (c=2, deadline=0.4).}
	\label{fig:exp-users}
\end{figure*}
In this experiment, as depicted in Figure \ref{fig:exp-deadline}, we vary the maximum tolerable deadline for the perception task, ranging from 0.4 seconds to 0.7 seconds, while keeping the number of users fixed at 40 and maintaining all other parameters constant. 
As the deadlines get extended, more users are permitted to offload to the same worker, as shown in Figure \ref{fig:exp-deadline-pi}, resulting in the fusion of more perceptions and a better coverage of the ROI. Consequently, all of the offloading schemes, GO, CPTO, Q-CPTO, and Q-CPTO-H, exhibit higher average detected awareness levels, as shown in both Figure \ref{fig:exp-deadline-awarness} and Figure \ref{fig:exp-deadline-sat}. This increase in awareness, however, comes at the cost of higher response delay, as more users share the same worker, leading to increased delay, as indicated in Figure \ref{fig:exp-deadline-delay}.

In our analysis, Q-CPTO and Q-CPTO-H effectively reduce the number of collaborating users by up to 30.4\% and 31.1\% when compared to GO, respectively, while outperforming it in terms of average detected awareness by up to 16.6\% and 17.6\%. This trend is also reflected in response delays, with Q-CPTO and Q-CPTO-H achieving 19.6\% and 27.1\% lower average response delay compared to GO. Moreover, both Q-CPTO and Q-CPTO-H achieve lower response delay compared to CPTO, with reductions of up to 10\% and 18.2\%, respectively, while achieving significantly higher ROI coverage of 36.2\% and 37.5\%. Additionally, they exhibit superior satisfaction ratios, surpassing GO by up to 14.3\% and 12.6\%, and outperforming CPTO by 30\% and 28\%. These results highlight the ability of the proposed schemes to efficiently offload users, reducing response delay compared to uniform offloading while achieving substantially higher ROI coverage. 

Furthermore, Q-CPTO-H demonstrates small performance gaps of up to 1.3\%, 2.9\%, 9.2\%, and 1.4\% in terms of average detected awareness, number of cooperating users, response delay, and satisfaction ratio, respectively, compared to Q-CPTO.
\begin{figure}[ht!] 
	\centering
	\includegraphics[width=3in]{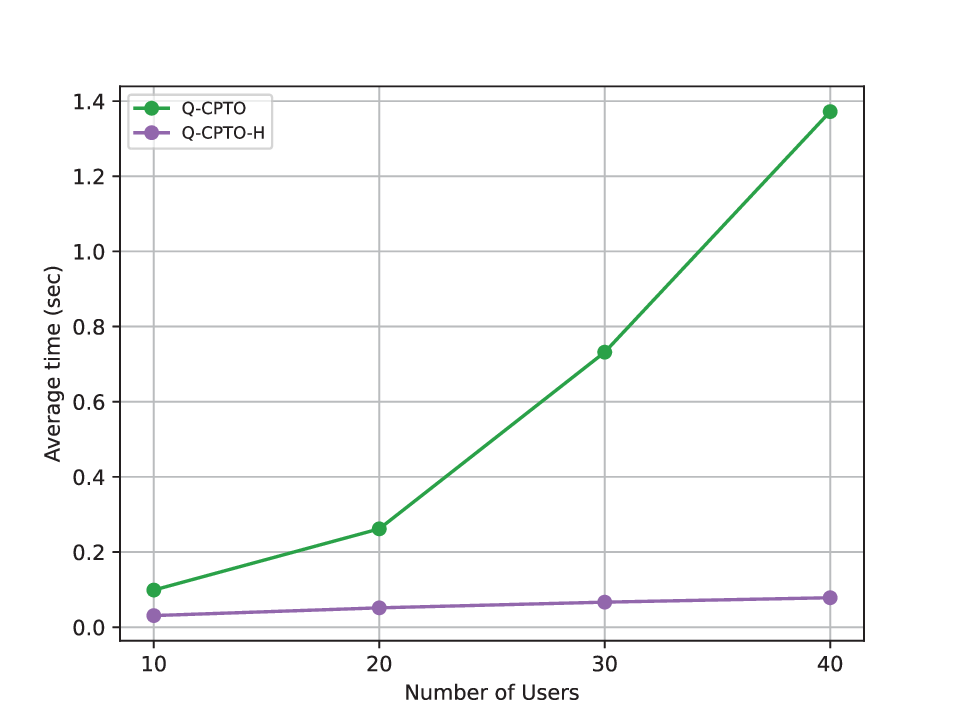}
	\caption{Run Time of Decision Making for Q-CPTO, Q-CPTO-H.}
	\label{fig:run-time}
\end{figure}
\subsubsection{Impact of varying number of users}~
In this comprehensive experiment, as illustrated in Figure \ref{fig:exp-users}, we vary the number of users in the system, ranging from 10 to 40, to conduct a comparative analysis of the proposed offloading schemes. 
As shown in Figure \ref{fig:exp-users-pi}, with an increase in the number of users, more users are directed to the same worker, resulting in the fusion of more perceptions and subsequently enhancing the detected awareness for all studied schemes, as indicated in Figure \ref{fig:exp-users-awarness} and Figure \ref{fig:exp-users-sat}. However, it is important to note that as the number of collaborating users increases, the response delay of the perception task also rises, as demonstrated in Figure \ref{fig:exp-users-delay}.

In this context, both quality-aware schemes, Q-CPTO and Q-CPTO-H, exhibit superior ROI coverage compared to GO by 13.6\% and 13.7\%, respectively, while employing a lower number of collaborating users. Specifically, Q-CPTO and Q-CPTO-H employ 25.7\% and 24.8\% fewer users compared to GO. This highlights the efficiency of these schemes in selectively offloading users, in line with our previous findings. Moreover, this efficient offloading results in lower response delays compared to GO, with both Q-CPTO and Q-CPTO-H achieving reductions of up to 13.5\% and 21.2\%, respectively. When compared to uniform offloading, both quality-aware schemes exhibit lower response delays of up to 6.6\% and 14.4\%, respectively, while achieving superior traffic awareness of up to 36.5\% and 36.7\%. Furthermore, both quality-aware schemes demonstrate higher satisfaction ratios compared to the studied schemes. Particularly, Q-CPTO and Q-CPTO-H achieve satisfaction ratio increases of up to 11.3\% and 10\% over GO, while outperforming CPTO by up to 30\% and 28.6\%, respectively. 

Additionally, Q-CPTO-H shows performance gaps of up to 0.3\%, 7.3\%, 8.8\%, and 1.1\% for average detected awareness, number of cooperating users, response delay, and awareness satisfaction ratio, respectively, compared to Q-CPTO. These results affirm the effectiveness of quality-aware offloading strategies across varying user counts, improving both awareness and response times while applying a selective approach for users required for collaboration.

\subsubsection{Time performance of offloading decision latency}~\\
In this final experiment, depicted in Figure \ref{fig:run-time}, we assess the scalability of the centralized SD-CSC by investigating how the number of users influences the time performance of the offloading decision scheme.
As evident in Figure \ref{fig:run-time}, with an increase in the number of users within the system, the computational complexity of running Q-CPTO rises accordingly. Given that Q-CPTO employs exhaustive optimization methods such as branch and bound, its time complexity increases significantly. However, it is noteworthy that Q-CPTO-H consistently maintains a much more efficient time performance, even when dealing with a high number of users in the system. Q-CPTO-H reduces the time required to reach offloading decisions by up to 83.5\%, demonstrating its  scalability and efficiency in handling larger user populations while maintaining swift decision-making processes.

%% file: 2e.conclusion.tex
\section{Conclusion and Future Work}
\label{sec:conclustions}

In this paper, we introduce the Quality-aware Cooperative Perception-based Task Offloading (Q-CPTO) scheme, a novel approach designed to enhance the cooperative perception (CP) capabilities of connected Autonomous Vehicles through selective offloading of perception aggregation tasks to Vehicular Edge Computing (VEC) workers. Q-CPTO leverages the diverse fields of view (FOVs) of vehicles, employing a shared interest utility function that predicts the vehicles' future trajectories and guides the selective offloading process.
Moreover, we propose Q-CPTO-H which employs an efficient heuristic based on reactive local search to reach sub-optimal solutions in polynomial time.
Our comprehensive comparative analysis, involving multiple baseline methods, demonstrates the effectiveness of our quality-aware schemes in reducing response delay by up to 14\% compared to a greedy offloading strategy while achieving superior ROI coverage, surpassing it by up to 20\%. Furthermore, Q-CPTO and Q-CPTO-H outperform CPTO by achieving up to 7\% lower response delay while significantly enhancing detected awareness by up to 44\%.

%% file: 3b.references.tex
\bibliographystyle{IEEEtran}
\bibliography{3c.bibfile}